# Impact of Local Energy Markets on the Distribution Systems: A Comprehensive Review

Viktorija Dudjak[1*], Diana Neves[2], Tarek Alskaif[3], Shafi Khadem[4], Alejandro Pena-Bello[5], Pietro Saggese[6], Benjamin Bowler[1], Merlinda Andoni[7], Marina Bertolini[8], Yue Zhou[9], Blanche Lormeteau[10], Mustafa A. Mustafa[11][12], Yingjie Wang[12], Christina Francis[13], Fairouz Zobiri[12], David Parra[5], Antonios Papaemmanouil[1]

1. Lucerne University of Applied Sciences and Arts, Switzerland
2. IN+ Centre for Innovation, Technology and Policy Research, Instituto Superior Técnico, Universidade de Lisboa, Portugal
3. Wageningen University and Research, Netherlands
4. International Energy Research Centre, Tyndall National Institute, Ireland
5. Energy Efficiency Group, Institute for Environmental Sciences and Forel Institute, University of Geneva, Switzerland
6. IMT School for Advanced Studies Lucca, Italy
7. Heriot-Watt University, United Kingdom
8. University of Padova, Italy
9. Cardiff University, United Kingdom
10. Université de Nantes, France
11. Unversity of Manchester, United Kingdom
12. KU Leuven, Belgium
13. University of Edinburgh, United Kingdom

**Abstract:**

In recent years extensive research has been conducted on the development of different models that enable energy trading between prosumers and consumers due to expected high integration of distributed energy resources. Some of the most researched mechanisms include Peer-to-Peer energy trading, Community Self-Consumption and Transactive Energy Models. Here, we present an exhaustive review of existing research around Local Energy Market integration into distribution systems, in particular benefits and challenges that the power grid can expect from integrating these models. We also present a detailed overview of methods that are used to integrate physical network constraints into the market mechanisms, their advantages, drawbacks, and scaling potential. We find that financial energy transactions do not directly reflect the physical energy flows imposed by the constraints of the installed electrical infrastructure. Therefore, ensuring the stable and reliable delivery of electricity, it is vital to understand the impact of these transactions on the grid infrastructure, e.g., impacts on the control, operation, and planning of electricity distribution systems.

**Keywords:** Local Energy Markets; Peer-to-Peer; Transactive Energy; Network Constraints; Impact; Distribution System Integration

1. Introduction

In June 2018, the European Union (EU) agreed a legal framework for prosumership, as part of the recast of the Renewable Energy Directive (RED II) [1]. This puts consumers in the centre of energy transition and introduces Citizen Energy Communities and Renewable Energy Communities, encouraging consumers to acquire ownership in distributed energy resources (DERs) and become prosumers - individuals who both consume and produce energy [2]. The framework is supporting the integration of DERs in the distribution network that can potentially provide services to the power systems [3] and calls for new energy business models.

[*] corresponding author: E-mail address: viktorija.dudjak@hslu.ch

DERs are defined as small or medium-sized resources, directly connected to the distribution network [4]. They include distributed generation, energy storage systems (ESS) and controllable loads, such as electric vehicles (EVs), heat pumps or demand response (DR). A high penetration of DERs can potentially be problematic for the stability and reliability of the distribution network and is expected to cause over-voltages, under-voltages and congestion [3], phase unbalance that may have negative impact on power quality [5], and unpredicted bi-directional power flows [6] for which the system was not originally designed. On the contrary, if managed intelligently, DERs could provide ancillary services to system operators through price-based incentives [7] as well as local system services to Distribution System Operator (DSO) to solve issues related to voltage regulation, power quality and distribution network congestion [8]. Beyond economic and technical aspects, other drivers that must be addressed by developed business models, are the social and environmental concerns, as well as privacy issues regarding the origin of energy, among households and business customers.

Increasing trends of DER deployment and grid digitalization [9] allow for the emergence of decentralized energy exchange paradigms to promote endogenous and local resources, increasing environmental benefits. One example of this is the Local Energy Market (LEM). Mengelkamp and Weindhardt [10] defined LEM as a socially close community of residential prosumers and consumers that have access to a joint market platform for trading locally produced electricity among each other. Such user-centric markets can be typified in Peer-to-Peer markets (P2P), Transactive Energy markets (TE), and Community Self Consumption (CSC). The Common denominator of the above mentioned models is that they use information and communication technology (ICT) for sustainable and efficient energy transactions [11]. In the context of P2P market platforms, the most used technologies are distributed ledger technologies, namely, blockchain [12].

P2P electricity trading is a business model, first proposed in 2007 [13], based on an interconnected platform that serves as an online marketplace where consumers and producers "meet" to trade electricity directly, without the need for an intermediary [12]. Since then, P2P electricity trading has risen in popularity within research as one of the possible paths to encourage power systems energy transition as it is expected to decrease participants' electricity bills by trading within peers and increase self-consumption of (surplus) locally produced renewable energy, in opposition of being totally supplied under the rules of a centralized retailer or market [14], [15]. In addition, it is claimed that P2P markets are fairer and more transparent [16]. P2P markets have been in the focus of an increasing number of pilot and demonstration projects in the recent years [17]–[35]. Different market structures can be implemented for P2P energy trading, for example centralized community-based markets, and distributed bilateral trading market [36].

A TE system is defined as a set of mechanisms that use economic-based instruments to achieve a dynamic balance between generation and consumption while considering operational constraints of the power system [37]. Within the TE system, DER generation and consumption can automatically negotiate their actions with each other using energy management systems and electronic market algorithms, allowing a dynamic balance of supply and demand [38]. Often TE is used interchangeably with P2P. However, TE represents a broad set of activities that includes much more than energy exchange transactions between peers [39].

CSC is a framework that supports the energy transition in the electricity sector by facilitating the collective sharing of renewable electricity generation assets within a community of prosumers, generally restricted to a neighbourhood, a district or an industrial consortium connected to the public network. It allows multiple end-users to benefit from shared distributed generation installations [40]. Such communities can be an actor in TE models or recognize each other as peers (similar to P2P models) and create a nested community-of-communities [41].

Electricity trading is different to other forms of exchange or trading of goods for two main reasons: (1) as opposed to other goods, electric energy cannot be stored economically and on a large-scale; and (2) electricity generation must match simultaneously electricity demand, considering that electricity delivery is implemented according to physics laws [42]. Customers are part of a power system, and in case of small customers, largely connected to a distribution network. Distribution network imposes technical constraints on the energy trading, and these constraints need to be represented in trading models in some way. While a certain schedule of DER and local consumption may be profitable from an economic perspective, these actions might violate current network constraints and cause reliability issues. One of the major challenges in implementation is to assure that network constraints are not violated during the energy trading [43], therefore constraints such as line, cable or transformer limitations and bus voltages should ideally be taken into consideration in the design of LEM models [44] [45]. P2P might also contribute to changes or relaxation of some of the constraints, or even force a redesign of the network [46]. Network constraints and integration issues of LEM are part of the physical LEM layer (layer 1), which together with ICT (layer 2), market (layer 3), economic (layer 4) and policy and regulation layer (layer 5) defines the high-level architecture of LEMs presented in Figure 1.

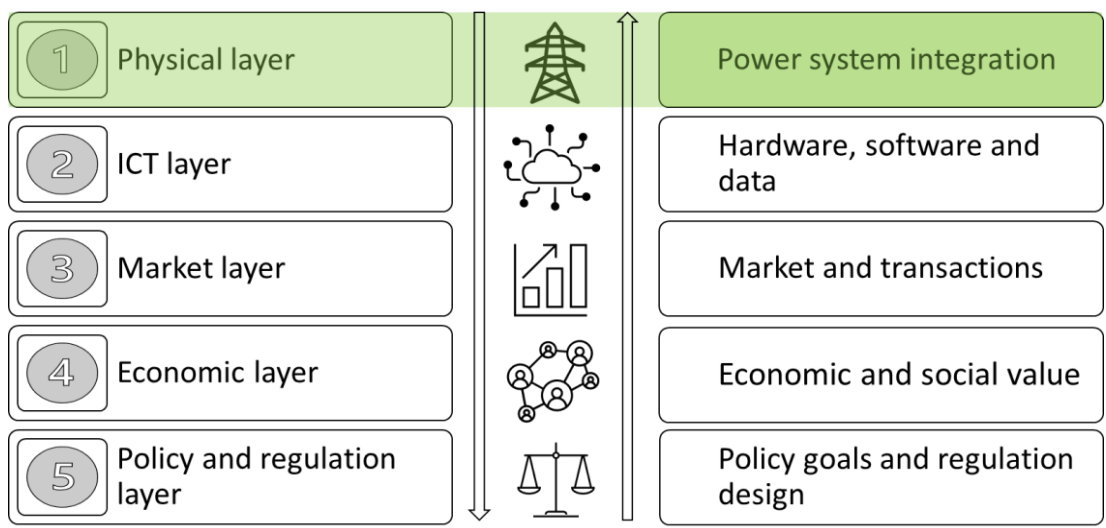

*Figure 1: Five layers architecture of LEMs*

In the recent years, a number of literature review publications has covered different aspects of LEMs, including P2P energy trading [47],[48] and blockchain technology implementation possibilities [49], [50]. Nevertheless, in LEM research and pilot projects to date, insufficient attention has been given to the physical layer issues and few articles, for example, Tushar et al. [45] refer to this topic in more detail, which leads us to conclusion that a more profound assessment of this issue should be taken in the literature. To bridge this knowledge gap, in this paper we focus on the integration and impact of LEM on power systems, by performing an in-depth and systematic literature review of the state-of-the-art, extending the review to classify the impact in specific technical characteristics of the distribution network. More specifically, the main contributions of this paper are to:

- Identify the impact of LEM operation on distribution network operation, planning and constraints;
- Provide an overview of commonly used methods to include the network constraints dimension in the LEM modelling;
- Identify knowledge gaps and open future research directions.

The structure of the paper is organized as follows. Section 2 provides an overview of the research methodology used to study the state-of-the-art in integration of LEM with power systems. In Section 3, the possible impact of LEM models on the distribution network are described. Section 4 provides an overview of methods to include physical grid parameters and Section 5 describes methods to allocate network fees and power losses using LEM transaction data. Finally, Section 6 discusses research gaps and future research directions and Section 7 concludes the paper.

## 2. Methodology

This section describes the literature review methodology. In this work we intend to deepen the impact of LEM on power systems, responding to the following research questions:

1. How can LEMs affect grid operation?
2. What are the possible problems and benefits of LEM operation?
3. How can self-consumption in the context of LEMs impact the grid?
4. What are the impacts of interaction between multiple LEMs?
5. What are the methods to include network constraints in LEM models?
6. What is the impact of LEMs on transmission network operation?

To define a paper selection metric, we used two review stages. First, the Web of Science (https://apps.webofknowledge.com/) databases were searched, with the databases being accessed during the period from July 2020 to January 2021, using following inclusion criteria:

*AB = (("peer to peer market" OR "peer-to-peer market" OR "P2P market" OR "local energy market" OR "local energy markets" OR "self consumption" OR "transactive energy market" OR "energy community") AND (distribution grid OR distribution network) AND (impact OR constraint)).*

All relevant papers were included, irrespective of publication date. Additionally, authors included papers that they considered valuable according to their expert knowledge. The selected papers then went through a first review for analysis against the inclusion criteria listed below:

1. The paper was written in English;
2. The paper concerned LEM (P2P, TE, CSC);
3. The paper included the impact of operation of LEM (P2P, TE, CSC) on power systems;
4. The paper was published in a peer-reviewed journal or presented at a conference.

As a final step, the list of references at the end of each reviewed paper were considered, and additional, relevant papers were extracted where they met inclusion criteria.

A detailed data extraction table was then created, allowing the resulting selected papers to be reviewed in line with the previously proposed research questions ($2^{nd}$ review).

The paper identification and selection process and corresponding results are shown in Figure 2.

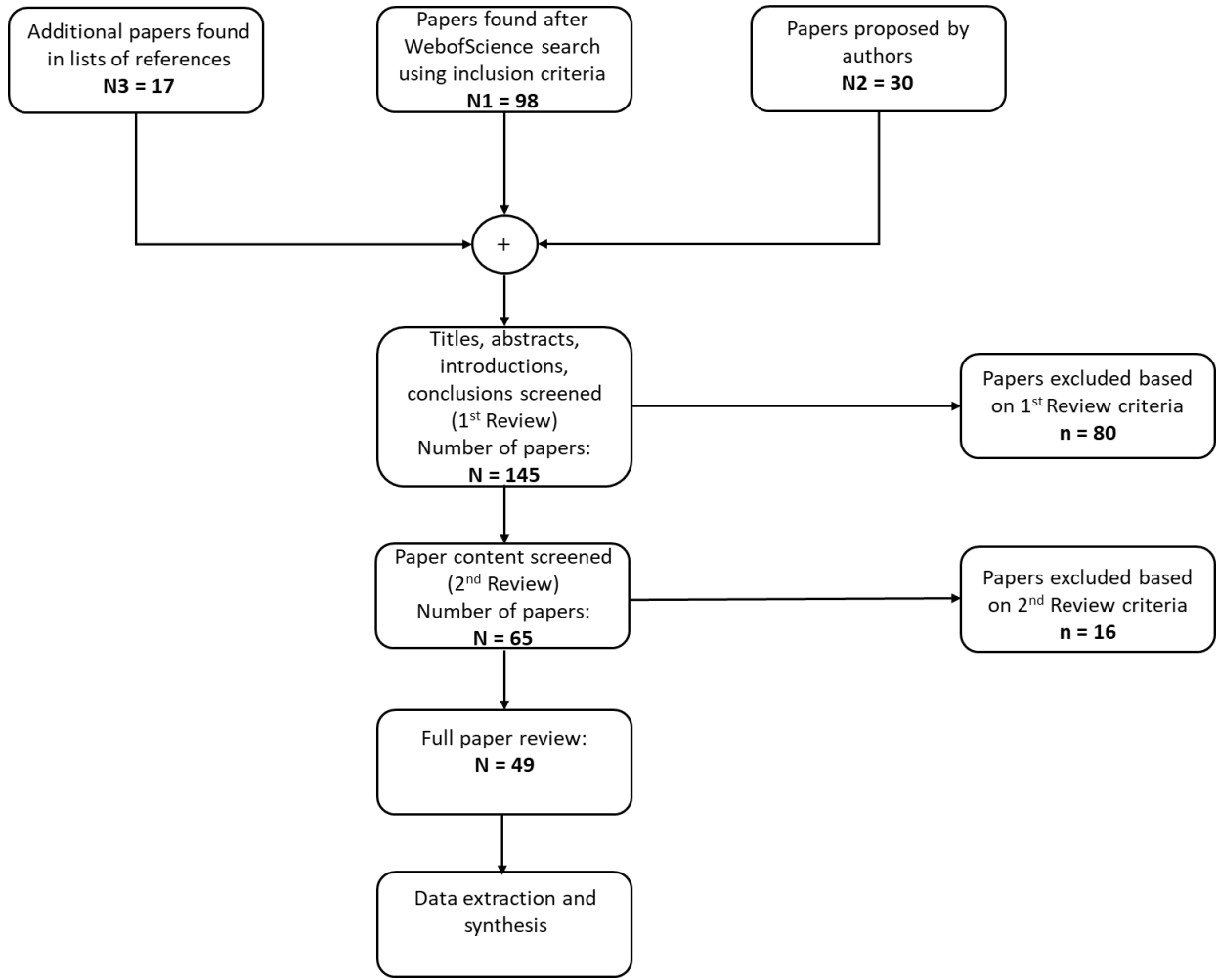

*Figure 2: Literature review process*

Out of 145 papers identified in the literature search, 65 papers passed the inclusion criteria (1st review). Of 65 papers that passed inclusion criteria, 49 papers went through the final review process. It is interesting to observe that although the inclusion criteria named different types of LEM plus reference to "distribution network" or "constraints", the papers that effectively directly address the LEM impacts on power systems were few. This indicates that although there is a common agreement that these kinds of market models impact power systems, the depth and terms of that impact is yet to be fully explored in literature.

Figure 3 shows distribution of papers that passed initial criteria review (1st review) and full review process (2nd review), according to their year of publication.

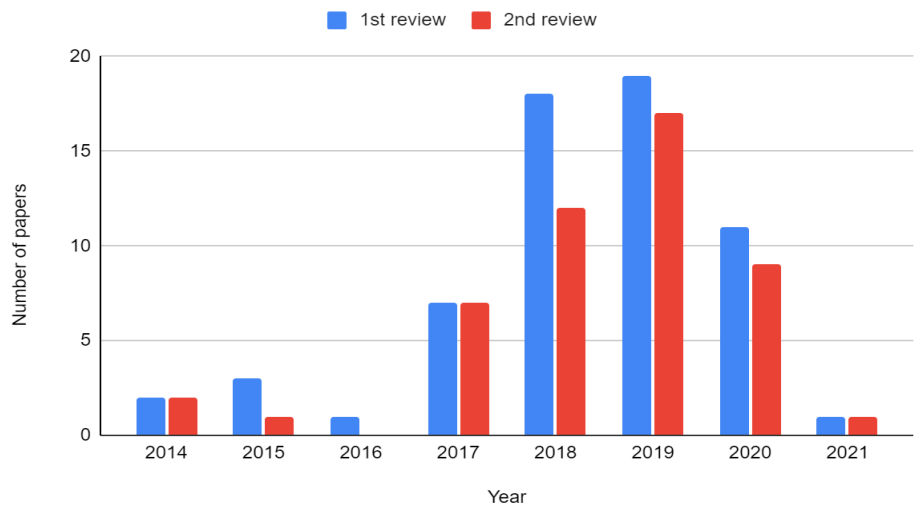

*Figure 3: Number of papers per year that passed 1st review (blue) and papers that went through 2nd review (red)*

The distribution of papers shows that the area of our search interest has been gaining in popularity since 2014. The number of published papers has been growing, especially in the period 2017 – 2019, with a smaller number of papers published in 2020. The number of papers published in 2021 is misleading as the literature search was finalized in January 2021.

During the review process, each analysed publication was assigned to at least one of the three following categories as shown in in Figure 4:

- Impact of LEM on the distribution network: such as in voltage variation, phase imbalance in LV network, system power peak, line congestion, cyber-attack vulnerability and distribution system planning; (Section 3)
- Methods to include physical grid constraints in market models, namely power equations, and network tariffs; (Section 4)
- Methods to calculate power losses and network tariffs that reflect trading flows between LEM actors (Section 5).

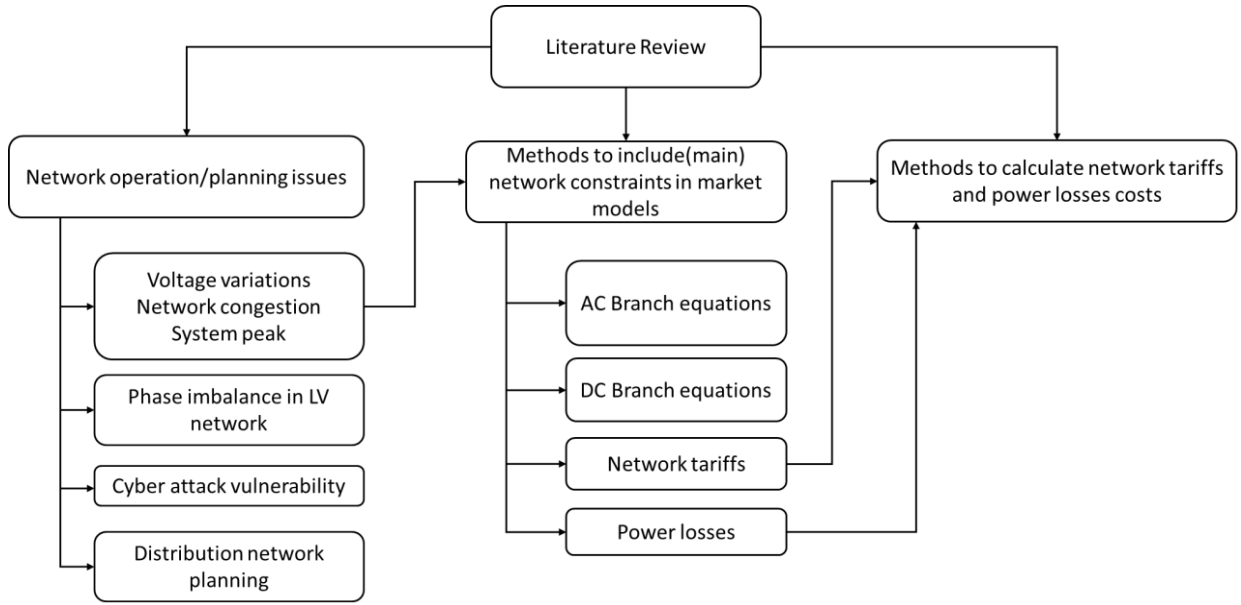

*Figure 4: Overview of paper categorization during research process*

However, regarding research questions number 4 and 6, we did not find scientific evidence on the studies analysed, worth reporting.

## 3. Impact of LEM models on the distribution network

Historically, distribution grids have been designed and operated in a centralized manner, with a unidirectional power flow in mind [44]. In this context, large generation units were responsible for power generation, which was injected into a high voltage (HV) transmission grid that had been adequately designed to transport large power quantities over long distances to load centres. In the contemporary system, the point of connection between upstream (transmission) and downstream (local, distribution) grid is usually a transformer (in a substation), after which power is often delivered to final customers using a radially operated, weakly meshed distribution grid. Radially operated feeders are designed to support worst-case peak loading expected across the feeder coming from the upstream grid. This planning approach is designed to sustain an 'N-1 redundancy criterion', a requirement to ensure quality and security of supply are maintained within the network. The criterion can lead to extensive capital investments with a high probability that the resulting network will be over dimensioned and only partially utilized. However, it is not clear if this design can cope with LEM trading volumes coming from DERs connected to the LV network.

LEMs have anticipated benefits, especially in the terms of more efficient grid utilization, reduction of exchanges of the LEM network (when defined on specific local area) with the main grid [51] due to local match of supply and demand. However, LEMs can also potentially create issues within the distribution network. Existing research primarily considers the impact of high penetration of DER on the low voltage (LV) and/or medium voltage (MV) grid when evaluating the impact of LEM models in distribution systems and how this impact can be mitigated. It identifies voltage variations, phase imbalance, impact on peak power and congestion, impact of LEMs on cyber-attack vulnerability, increased complexity of distribution network planning and increased complexity of control. Table 1 summarizes the expected impact of LEMs on the distribution network, as identified during the literature review process.

*Table 1: Summary of detected technical impact of LEMs*

| Technical impact | Impact | Reference | Major impact driver |
|---|---|---|---|
| **Voltage variation** | High voltage | [52] | Scenario with high rooftop PV penetration level |
| | | [53] | Scenarios without including prosumers' trading strategies |
| | Minimization of voltage drops | [53] | Affected by prosumers' trading strategies |
| | | [54] | Once LEM mechanism is introduced, voltage issues are removed |
| | No specific effect | [55] | - |
| **Phase imbalance in LV network** | Imbalance across different phases | [56], [57], [58] | Affected by prosumers' trading behaviour Introduction of control mechanism does not have negative effect on market mechanism outcome [58] |
| **System power peak** | Increased system power peak | [59] | Scenarios without including prosumers' trading strategies |
| | Reduced power peak | [51] | Reduction of peak load in scenario with LEM when compared to base case (no LEM) |
| | | [56] | Battery installation in LEM has high impact on peak reduction |
| | | [60] | Network capacity tariff has higher impact on peak power reduction |
| | | [61] | Impacted by prosumers strategies and market mechanism |

| Technical impact | Impact | Reference | Major impact driver |
|---|---|---|---|
| | | [62] | Impacted by flexibility market shifting demand |
| | | [63] | Impacted by inclusion of physical network constraint |
| **Line congestion** | Reduced line congestion | [59] | Scenario with uniform pricing mechanism |
| | Increased line congestion | [59] | Scenario with heterogeneous pricing mechanism |
| | | [64] | Scenarios without network fees and with unique (constant) network fee |
| **Cyber-attack vulnerability** | Reduced vulnerability | [59] | Distributed management approach |
| | Increased vulnerability | [65] | Increased vulnerability on electricity prices attacks |
| **Distribution system planning** | Reduced investment needs | [66], [67] | - |

As seen in Table 1, different studies show diverging results. In our opinion, one of the reasons for diverging results is the scenario-based design of current studies. Studies are normally performed on synthetic test cases, with assumptions about possible prosumers' trading strategies and defining differently market models, therefore deriving results that are heavily scenario specific.

Table 2 shows a summary of most used test cases in the reviewed literature, according to the type of data used, as well as detailed simulation data.

*Table 2: Summary of use cases in reviewed literature*

| Test case | Voltage level | Simulation data | |
|---|---|---|---|
| | | **Synthetic** | **Smart Meter / Real system data/ Daily representative curves** |
| IEEE LV European Feeder | Low voltage | [53], [55], [68] | [62] |
| IEEE 9 bus test system | Medium voltage | [64] | |
| IEEE 13 bus test system | | [69] | |
| IEEE 14 bus test system | | [59] | |
| IEEE 37 bus test system | | [54] | |
| IEEE 39 bus test system | | [64], [70] | |
| IEEE 69 bus test system | Medium voltage | [24], [65], [71]–[73] | |
| IEEE 123 bus test system | | [54], [69] | |
| Various non-standard LV test systems | Low voltage | [43], [51], [56]–[58], [71], [74], [75] | [63] |
| LV test systems based on real system characteristics | Low voltage | | [76], [77] |

In the following subsections, we present analysed research work and studies on the impact of LEM models on distribution network infrastructure, according to the category of impact, as presented in Table 1. We further analyse how that impact can be mitigated by modelling different prosumer behaviour and/or market design mechanisms, the main conclusions they bring as well as identified gaps for future research.

*Voltage variations*

Much of the existing research identifies voltage variations as the biggest possible challenge arising from LEM models. Voltage fluctuations are systemic variations of the voltage, the magnitude of which does not normally exceed specified voltage ranges (i.e. 0.9 to 1.1 p.u.) [78]. The main drivers for voltage variations are high DER penetration and the number of simultaneous energy transactions between prosumers. Azim et al. [52] reveal that simultaneous P2P transactions can raise the bus voltages beyond the limits defined in the grid code. Therefore, P2P trading inside a single feeder has the potential to cause over-voltage in the network. Conversely, if photovoltaic (PV) inverters are equipped with voltage controllers, many of these transactions will be curtailed for voltage regulation and non-dispatchable PV is the most common DER at household level since other DERs like wind, geothermal, biogas are location-specific [79]. The impact of P2P trading on voltage variations has also been studied by Herencic et al. in [53]. The study shows that voltage levels, as well as power flows, are primarily affected by prosumers' strategies for demand. They argue that effects of energy trading on voltage levels primarily depend on the level of power flows coming from and/or going to the upstream grid and conclude that improvement of local electricity supply-demand balancing behind the substation, driven by change in pattern of local demand, leads to minimization of voltage drops and increases voltage levels. A positive impact on voltage variations is further presented in [54] by including a power flow equations and voltage constraints optimization model of TE (more details on constraint modelling is given in Section 4). Studies performed show that, without TE and with observed PV penetration, overvoltage violations occur concurrently with peak PV generation in the system and undervoltage occurs when peak load occurs. When TE is introduced, all voltage problems are removed, since the market mechanism also includes network constraint optimization. Hayes et al. [55] indicate that a moderate level of P2P energy trading (more precisely, at a level that does not increase peak demand of the system) should not have a significant impact on network operational performance in terms of phase voltage imbalance and voltage profiles. Nousdilis et al. [68] investigated to what extent the self-consumption rate (SCR) of prosumers in an LV feeder can affect the voltage quality. The results show that consumers must effectively maintain their average monthly self-consumption rate above a certain system defined value depending on the quality of the network to which they are connected. Jhala et al. [72] have developed a new analytical method for voltage sensitivity analysis that allows for stochastic analysis of change in grid voltage due to change in consumer behaviour and to derive probability distribution of voltage change on buses due to random behaviour multiple active consumers, for both fixed [72] and the spatially random [71] distribution of active consumers.. In [73] they developed a data-driven method that mitigates voltage violation by taking a control action before actual voltage violation happened. To date, the method has been developed and tested for only single-phase systems.

*Phase imbalance*

Phase unbalance involves both voltage unbalance and current unbalance [80]. IEC defined current unbalance factor as the ratio of the negative sequence component to the positive sequence component [81][82]. As the consequence of voltage and current unbalance, the power values on the three phases are also unbalanced. Most LEM studies assume balance between phases and do not consider the phases to which households are connected. Network imbalance between phases can lead to bigger voltage rises and higher losses. Horta et al. [57] presented a method to minimize the negative impact of market participants, which are considered to have the highest impact on voltage unbalance due to their DER installation. This is ensured by dynamic phase switching by the system operator. The paper presents results of a simulation that shows dynamic phase switching does not have a negative impact on the outcome of the LEM (market mechanism explained in [56]) and can effectively increase the capacity of the distribution grid for hosting renewable energy. Further, in [58], they included a real-time control mechanism that copes with forecast errors by driving households towards a final exchange with the grid that benefits the prosumer and respects the DSO's quality of supply requirements, in particular voltage deviations and current intensities along the feeder. Hayes et al. [55] showed that Phase Voltage Unbalance Rate (PVUR), the maximum voltage deviation from the average phase voltage as a percentage of the average phase voltage slightly reduced in the P2P case (as compared to the base case without trading).

*Increased power peak and congestion*

LEMs have the potential to increase in DERs in distribution networks and so may subsequently cause increased congestion in the system due to the absence of matching generation and available transmission infrastructure hosting capacity [83]. Congestion is also caused by unexpected eventualities such as generation outages, unexpected escalation of load demand, and equipment failure [84]. Le Cadre et al. [59] simulated impact of different price distributions (uniform, heterogeneous, symmetric, and local trade preferences with uniform prices) on congestion in the network and concluded that price development mechanisms impact the outcome of the LEM and can cause congestion. Energy quantities traded in case of heterogeneous prices are much larger and almost half of the lines are congested, whereas some lines are almost unused in case of uniform market prices. This leads us to the conclusion that network impact of LEMs is heavily dependent on market and pricing design. Besides these two mechanisms, network tariffs can also have an impact on changing prosumers' behaviour and subsequently impact on network (more details on network tariffs modelling are given in Section 5). Almenning at al. [60] studied how network tariffs and P2P trading affect the energy import management of a small neighbourhood that is able to trade energy locally as well as utilize several different flexible loads. Two network tariff structures were modelled (capacity and energy based) on two levels (neighbourhood and consumer). For the consumer level, all consumers worked individually and were unaffected by the operation of other consumers. Results show decreased power peak by 11% and 7% if considering a consumer level and neighbourhood level, respectively. A capacity subscription tariff (instead of an energy tariff) registered the lower grid imports in the neighbourhood. Tushar et al. [61] also proposed a P2P energy trading scheme that could help a centralized power system to reduce the total electricity demand of its customers at the peak hour. Morstyn et al. [62] studied how the DSO could manage overall distribution peak demand by obtaining flexibility from aggregators and prosumers with small-scale flexible energy resources. These types of flexibility markets could also be integrated into future P2P electricity markets. One of DSO management options in reducing local grid peaks is also by integrating storage devices (either community-based or local) together with energy management systems [85]–[88].

*Vulnerability to cyber attacks*

Since LEM models rely on the data coming from smart meter devices, cyber security attacks pose a risk to the distribution grid operation [89], although this risk is not exclusively of LEM or P2P architectures [90]. Jhala et al. [65] investigated the impact of a false data injection attack by simulating an attack on demand data and an attack on electricity price signals. Results show that the impact of an attack on electricity prices is more severe than an attack on electricity demand since the attack on electricity prices requires manipulation of only one parameter. Le Cadre et al. [59] note that in case of failure or if one node is attacked, the power system can still rely on the other nodes as the information and decisions are not optimized by a single central entity. This could increase resilience when compared to a centralised management approach.

*Distribution network planning*

The planning of distribution networks in the LEM network environment has not yet been widely studied in the identified literature, although planning frameworks to incorporate flexibility into the planning process have been proposed [91], as well as methodologies for joint planning and operation of distribution networks [92]. Delarestaghi et.al. [66] studied how the inclusion of a P2P market affects the investment plan of different stakeholders in the distribution network. The study showed that the deployment of a P2P market results in less energy purchased during peak hours, which in turn means less power passed through the substation and feeders, helping to prevent the utility from unnecessary investment. Difficulties associated with network planning in the context of DERs do not come from the market itself, but rather from the risk of consumers disconnecting from the network, increased costs of facilities and equipment common to the network for consumers remaining connected, increased operating costs and increase in electricity prices due to climate policies. Delarestaghi et al. [67] developed a novel distribution planning framework that uses scenario-based investment planning approach by clustering historical data (energy wholesale prices, loads and PV generation) into several representative day clusters and solved the optimization problem using mixed integer second-order cone programming (MISOCP). The paper showed that, for scenarios where neighbourhood energy trading is allowed, the total cost of electrification decreases, while end-users' investment in batteries and PV units increases.

*Control mechanisms*

With the grid under increasing stress because of growing reliance on electricity and the introduction of DERs, the role of the DSOs in controlling quality of supply within the allowed limits is increasingly challenging. Reinforcing or replacing parts of the system is expensive and time consuming. A possible solution is to actively use the active and reactive power control capabilities of those DERs to keep the voltage within limits. To cope with problems coming from increased DERs penetration and integration of P2P markets in the network, the SmarTest project [93] investigated different methods to deliver P2P schemes, including distributed grid control, multi-agent systems, coordination across different control algorithms, use of power electronic devices and decentralised voltage control algorithms. Almasalma et al. [94] developed a voltage control algorithm that regulates the voltage within allowed limits. The approach is based on dual decomposition theory, linearization of the distribution network around its operating points and P2P communication and its experimental validation was presented in [95]. The results show that distributed voltage control systems can provide satisfactory regulation of the voltage profiles and could be an effective alternative to centralized approaches. The proposed P2P system could help in delivering easier access to prosumers' flexible supply and demand by making their active participation in the grid possible, and subsequently making LEM easier to integrate in the existing system.

4. **Methods to include physical grid constraints in market models**

Methods to include a consideration of the physical grid layer alongside the market layer vary to a great extent in literature. In the previous section we have mentioned that the physical impact of LEM models can vary greatly depending on whether network constraints are implemented in the market model. In this section, we describe methods for including network constraints in market mechanisms in more detail based on our literature research. We cover in detail the integration of branch flow equations into market mechanism models to accurately reflect grid constraints and/or give dynamic price signals to the prosumers to adapt their behaviour in a way to comply with network constraints. Additionally, power losses and network fees can also be used as signals to include network constraints, due to LEMs and accompanying ICT infrastructure and we summarize them in Table 3 alongside branch equations. Since network tariffs and power losses costs are not primarily used only to mimic network constrains, but to recover network costs, they are covered separately and in more detail in Section 5.

*Table 3: Overview of selected research that considers grid constraints in market mechanism clearing algorithm*

| Reference | Network constraints in market clearing algorithm | Optimal power flow calculation |
|---|---|---|
| Guerrero et al. [43] | - Voltage Sensitivity Coefficients<br>- Power Transfer Distribution Factors<br>- Loss sensitivity Factors | No |
| Azim et al. [52] | Yes/No<br>- power flow calculated after the market solution is obtained | No |

| Reference | Network constraints in market clearing algorithm | Optimal power flow calculation |
|---|---|---|
| Munsing et al. [96] | - branch flow equations | Yes – decentralized OPF |
| Li et al. [54] | - branch flow equations | Yes - decentralized OPF |
| Wang et al. [97] | - DC branch flow equations approximation | Yes |
| Qin et al. [98] | - DC branch flow equations approximation | No |
| Masood et al. [69] | - DC branch flow equations approximation | No |
| AlSkaif et al. [99] | - AC branch flow equations | Yes<br>AC OPF |
| Van Leeuwen et al. [63] | AC branch flow equations | Yes<br>AC OPF |
| Xu et al. [100] | - matching of supply and demand with minimum power transmission losses | No |
| Guerrero et al. [70] | - Voltage Sensitivity Coefficients<br>- Power Transfer Distribution Factors<br>- Loss sensitivity Factors | No |
| Baroche et al. [64] | - in form of network fees | No |
| Moret et al. [24] | - the form of spatial and temporal varying network fees | No |
| Zhong et al. [101] | branch flow equations | No |

*Branch flow equations*

Branch flow equations represent constraints imposed by power flows on radial distributions systems by substituting conventional AC power flow equations. They were introduced first by Baran and Wu [102] to model power flows in a steady state in a balanced single-phase distribution network.

We represent here an abridged version presented by Li et al. in [103]. We consider a radial distribution network that consists of a set of $N$ buses and a set of $E$ of distribution lines connecting these buses. We index the buses in $N$ by $I = 0, 1, ..., n$ and denote a line in $E$ by the pair $(i,j)$ of buses it connects. Bus 0 represents the substation and other buses in $N$ represent branch buses. For each line $(i,j) \in E$, let $I_{ij}$ be the complex current flowing from buses $i$ to $j$, $z_{ij} = r_{ij} + ix_{ij}$ the impedance on line $(i,j$, and $S_{ij} = P_{ij} + iQ_{ij}$ the complex power flowing from buses $i$ to bus $j$. On each bus $i \in N$, let $V_i$ be the complex voltage and $s_i$ be the complex net load, i.e., the consumption minus generation. As customary, we assume that the complex voltage $V_0$ on the substation bus is given. The branch flow model is given with the following set of equations:

$$p_j = P_{ij} - r_{ij}l_{ij} - \sum_{k:(j,k)\in E} P_{jk}, \qquad j = 1, \ldots, n \tag{1}$$

$$q_j = Q_{ij} - x_{ij}l_{ij} - \sum_{kk:(j,k)\in E} Q_{jk}, \qquad j = 1, \ldots, n \tag{2}$$

$$v_j = v_i - 2(r_{ij}P_{ij} + x_{ij}Q_{ij}) + (r_{ij}^2 x_{ij}^2)l_{ij}, \quad (i,j) \in E \tag{3}$$

$$l_{ij} = \frac{P_{ij}^2 + Q_{ij}^2}{v_{ij}}, \quad (i,j) \in E \tag{4}$$

where $l_{ij} \coloneqq |I_{ij}|^2$, $v_i \coloneqq |V_i|^2$, and $p_i$ and $q_i$ are the real and reactive net loads at node $i$. Equations (1) – (4) define a system of equations in the variables $((P,Q,l,v) \coloneqq (P_{ij}, Q_{ij}, l_{ij}, (i,j) \in E, v_i, i = 1, \ldots, n)$, which do not include phase angles of voltages and currents. Given an $(P, Q, l, v)$. These phase angles can be uniquely determined for a radial network.

In the next two sections we present current research in the area of LEMs that uses branch equations as a tool to include network constraints in the market mechanism design.

*AC branch flow equations*

Munsing et al. [96] propose an architecture for P2P energy markets to guarantee that operational constraints are respected, and payments are fairly rendered. The network was modelled as an undirected radial graph and power flow constraints formed a non-convex set. They used Alternating Direction Method of Multipliers (ADMM) to decompose the convex optimization problems resulting from the network and DERs' constraints. The work assumed that each party in the system had full knowledge of network topology in the system. Wang et al. [97] also included branch flow equations as network constraints to schedule DERs in an optimal way by solving the Optimal Power Flow (OPF). Further applications of OPF can be found in [54] which showed a positive impact on solving voltage variations (c.f. Section 3), as well as in [63] where the AC OPF problem was combined with a bilateral trading mechanism in a single optimization problem. It led to a fully decentralized algorithm that achieved maximum total social welfare by minimizing both grid import costs and trading costs for every agent separately and in parallel while respecting global grid constraints and balancing supply and demand. The model was tested with dataset from a real prosumer community in Amsterdam and results showed that inclusion of physical network constraints in the optimization problem meant the algorithm would avoid using the grid excessively during peak hours, not just because of cost incentives, but also because of possible congestion issues.

*DC branch flow equations*

Qin et al. [98] proposed linearizer DC approximation of the AC power flow equations. Constraints were modelled as capacity constraints and equality constraints over the entire network (demand equal to supply). The system operator ensured that network constraints were not violated by curtailing trades if network constraints were violated, and by publishing information about the network state to guide participants regarding how subsequent trades could avoid overloading congested lines. DC power flows were also included in optimization problems in [69] and [104] for an interaction between the DSO, TE market operator and aggregators that represent interests of flexible customers (e.g. EV owners). In the first stage, the aggregator collected the charging requirement of an individual EV. Based on these requirements an initial aggregated charging schedule of EVs was created and an energy profile was provided to the DSO. In the second stage, if the flexibility call was activated, the aggregators accumulated the available flexibility from consumers to offer bids in the form of flexibility profiles with the information about EVs that will refrain from charging. The study case showed that by incorporating network constraints in the bidding optimization problem, the solution was technically much more effective as it led to the activation of only technically feasible bids. The model was tested for a larger test network for scaling purposes and showed that it could be solved more quickly as a result of it being based on linear programming. Decentralized ADMM-based OPF on a private blockchain-smart contracts platform has been tested in [99]. Smart contracts could be expanded to allow trading mechanisms, although the study does not assume any trading between different households.

*Post market-clearing constraints*

Additional studies [98], [105] propose market models without network constraints, but in order to ensure that transactions do not cause violations, at a certain point of a time, the DSO collects all the contracts, and rejects those that cause violation.

## 5. Methods to calculate network tariffs and power losses that reflect trading flows

Currently, the main method for recovering distribution network costs is through network usage fees. As described in a report by the European Commission [106], the majority of distribution grid tariffs in Europe consist of volumetric charges (i.e. €/kWh). In a traditional setting, consumers connected to the distribution network are not able to react strongly to price signals and volumetric tariffs are only slightly cost-reflective. Higher penetration levels of DERs as well as introduction of LEMs at the consumer-side, are challenging the traditional use of volumetric network charges. Specifically, volumetric charges with net-metering, implying that a consumer will be charged for the net consumption from the grid over a certain period (e.g. month), are deemed inadequate with the massive deployment of solar PV[107]. In the context of LEMs, the objective of the network fees allocation could be for system operators to achieve cost recovery but also to reduce congestion risks (i.e., to influence prosumers to behave in a certain way).

The technological advances of LEMs give an opportunity for development of new allocation methodologies of power losses. There is a possibility to assign power losses to every transaction in the system, contrary to prevailing approach of evaluating power losses in the system by estimating them at the highest demand (using some loss estimation method) and applying the loss factor to predict the total energy losses [108]. Besides allocation, similar to network fees, power losses can also be used to mimic network constraints in the system as introduced earlier in Section 4.

Table 4 summarizes the research studies and methods used to allocate network fees and power losses in the system as well as where those methods were used as network constraint or price signals in LEM market models.

*Table 4: Overview of methods to calculate network tariffs and power losses that reflect trading flows*

| Reference | Allocation of network fees method | Allocation of power losses method | Included in market algorithm as constraint |
|---|---|---|---|
| Guerrero et al. [43] | PTSFs values | PLSFs, together with VSCs | Yes |
| Lilla et al. [74] | No | Proportionally attributed to the transactions that create flows in branch | Yes |
| Guerrero et al. [70] | PTSFs values | PLSFs, together with VSCs | Yes |
| Baroche et al. [64] | Exogenous costs (unique unit fee, distance unit fee, uniform zonal unit fee). | No | Yes |
| Paudel et al. [109] | Power transfer distribution factor | Yes | Yes, if factors published in advance |
| Moret et al. [24] | Relative transaction cost between energy communities | No | Yes |
| Zhong et al. [101] | Network usage tariff with defined upper and lower limit | No | Yes |
| Di Silvestre et al. [77] | No | Proportional Sharing Rule (PSR) index | No |
| Nikolaidis et al. [110] | No | Graph-based framework (3 phase) | No |

*Network fees*

Guerrero et al. [43], [70] proposed a methodology to assess the impact of P2P transactions based on voltage sensitivity factor (VSC), power transfer sensitivity factor (PTSF) and power loss sensitivity factor (PLSF). VSC was used to calculate voltage variations leading to transactions not being allowed where they caused voltage issues in the network. PTSFs values were proposed to assign congestion charges: agents paid charges for using a physical network, and this could also be used to estimate the congestion in the lines. PLSFs, together with VSCs, were used to calculate costs associated with losses caused by each transaction. Simulation results showed that the proposed method reduced the energy cost of the users and achieved the local balance between generation and demand of households without violating the technical constraints. Baroche et al. [64] tested three incentive frameworks in a form of exogenous costs (unique unit fee, distance unit fee, uniform zonal unit fee). The distance unit fee showed the ability to limit the stress put on the physical grid by the market. On the downside, the approach may lead to inefficient or unfeasible solutions when network charges are not chosen wisely.

Similarly, Paudel et al. [109] proposed a method to calculate network fees based on power transfer distribution factor. The network owner provided the charging rate for network utilization in advance before the P2P negotiation started. The network owner considered the capital cost recovery, cost of maintenance and modernization of power lines, taxes, and policies, etc. to decide the rate for the network utilization. Approximated losses were also considered. Moret and Pinson [24] investigated additional costs that mimic network constraints when an energy collective is formed by prosumers from different neighbourhoods. Flow was defined for each line connecting the neighbourhoods and geographical differentiation was included as a relative transaction cost. This formulation allowed representation of technical constraints, typical of power flow analysis, in the form of spatial and temporal varying grid tariffs. Zhong et al. [101] proposed a cooperative energy market model where buyers and sellers trade energy in a P2P manner and pay network tariff to the network operator. A network usage tariff that is too high discourages buyers and sellers from P2P energy trading, while a network usage tariff that is too low discourages the network operator from providing P2P power delivery services.

*Power losses costs*

Allocating power losses cost for each transaction between prosumers in LEM is a complex problem since the missing link between virtual and physical transactions makes correct power losses allocation difficult. Di Silvestre et al. [77] proposed a Proportional Sharing Rule (PSR) index that gives a more accurate evaluation of the power losses to be associated with the energy transaction between a specific couple generator/load. Nikolaidis at al. [110] proposed a graph-based framework for allocating power losses in 3-phase 4-wire distribution networks among the P2P contracts or energy communities. Each transaction was not only defined by the transaction path between nodes (that can be connected to different phases), but also with the "mirrored" path on the neutral layer. Results show that simplifying assumptions in terms of net demand unbalances at the LV level may introduce significant error in the calculated losses and their allocation. Omitting the influence of neutral flows on total losses may introduce non-negligible errors in the loss allocation process. Xu et al. [100] proposed a novel discounted min-consensus algorithm to discover the optimal electric power-trading route with minimal power losses in DC microgrids and avoid congestions in the grid. It considered network constraints in the form of power losses in power lines. An advantage of this approach is that it requires only local and neighbourhood information for each agent, without the knowledge of the system parameters. Lilla et al. [74] presented a method of day-ahead scheduling of LEM using ADMM, previously developed by Orozco et al. [75]. The goal of optimization is to minimise energy procurement costs of the community considering power loss in the internal LV network by allocating internal network losses to various power transactions between two prosumers or between a prosumer and the utility grid. The results confirm that, in the considered LEM framework, each prosumer achieves a reduction in costs or increases revenues by participating in the LEM compared to the case in which it can only transact with an external energy provider.

## 6. Discussion

In general, it was found that LEM research is very transdisciplinary, making it hard to decouple the impact on power systems from market model design or existing policy and regulation frameworks. Our results therefore lead us to conclude that research in this area is still limited by existing market and policy frameworks, presenting an opportunity for further research.

*Research design of previous case studies*

It was difficult to fully draw systematic conclusions due to the research design associated with previous work. Previous research was restricted due to limited or non-existent historical trading data in LEMs, meaning that research only had access to simulation data on future prosumers' behaviour. Prior work is therefore scenario-based and dependent on model assumptions for future prosumers' trading strategies. Prosumers' trading strategies were often modelled using mathematical optimization methods that might not unambiguously translate to reality once LEMs are implemented in the real distribution system. This makes it challenging to study their impact on distribution network operation. In addition, at times the assumptions for simulations presented in literature were contradictory, leading to opposing conclusions in relation to certain topics.

*Volage and congestion problems and benefits*

A large number of researchers anticipated voltage violations and congestion as being an important impact of LEMs. There appeared to be consensus that prosumer strategies for demand affect voltage levels, but the extent of the impact, and whether it is positive or negative, depends on the market mechanism employed. If a mechanism is employed that does not increase the peak demand of the system, there will not be a significant impact on network performance in relation to voltage imbalance and voltage quality. It should be noted that most studies focused on designing market models, control mechanisms and participant models that had a positive impact on voltage from the technical point of view, meaning negative consequences were avoided by design and so not observed in the results.

*Phase imbalance*

For phase imbalances, most studies assumed balance between phases and did not the phases to which households were connected. Network imbalance between phases can lead to higher voltage rises and losses, therefore future studies should investigate the effects of LEMs on phase imbalance issues in more detail, especially considering that most of the use cases are designed with prosumers connected to LV networks in mind.

*Network control solutions*

To realise benefits in practice, the system requires improved dynamic and decentralized network control solutions as well as active operation and planning from the DSO side. We believe the role of digitalization is crucial to enable that. Once digitized, systems become more efficient and responsive, which allows innovative solutions, including DERs and LEMs to integrate more rapidly. On the other hand, it gives DSOs an opportunity to increase their reliability, efficiency and customer engagement.

*Methods to include physical network constraints in market models and the role of DSO*

The review also reported methods to include physical grid constraints in market models, and the associated role of the DSO. While many OPF methods for clearing markets seem to be promising, as well as branch equations solutions to include network constraints in market mechanisms, they require either a centralized approach where central entities have infrastructure knowledge, or a distributed approach, where each prosumer needs to have infrastructure knowledge of their immediate neighbourhood. This leads to the conclusion that the DSO needs to be involved to a great extent in the LEM mechanism development and decision making. It is therefore of great importance to study how to integrate DSO or a central entity within the LEM in case of a centralized marketplace design. In case of distributed marketplace design, information sharing and responsibilities between involved actors need to be properly defined, especially when it concerns critical infrastructure information.

*Network tariffs as network constraints signals*

Research identified that, instead of branch equations more dynamic network tariffs, when compared to current practice, could be a signal to prosumers to change unwanted behaviour that led to undesirable consequences in local networks. Tariffs need to be designed in a smart way to efficiently recover network costs because they could lead to undesired effects, for example an inability of the DSO to recover its operational costs.

*Impact on the transmission network operation*

The literature review did not identify research works that studied the impact of LEMs on transmission system level operation and planning. Most of the study cases identified in the literature were based on small test cases (in terms of number of prosumers, peak power, feeder and location). The impact of LEMs could become a challenge to the transmission network in case of a high number of LEMs, so future research should cover LEMs providing services to transmission networks, reversible flows issues, network tariff redesign (i.e., to cover TSO cost recovery), and how the role of transmission network changes once a significant amount of DERs are connected to the distribution network and supplying LV customers almost exclusively through LEMs.

## 7. Conclusion

We conducted an extensive literature review to identify and discuss the network impact of integrating LEM models in the distribution network. The intention of the research was to identify how LEMs affect grid operation, the possible problems and benefits of LEMs, the impact of self-consumption on the grid in the context of LEMs, the impacts of interaction between multiple LEMs, the methods used to include network constraints in LEM models, and the impact of LEMs on transmission network operation. Of 145 papers identified on the topic, 49 papers were fully reviewed according to the inclusion criteria defined in Section 2 . No papers were identified that considered the interaction between multiple LEMs or the impact of LEMs on transmission network operation. Research on remaining topics was synthesised according to 3 topics of interest First, we categorized papers that deal with the impact of LEM on the distribution network (in relation to voltage variations, phase imbalance, power peaks, congestion, vulnerability to cyber attacks, network planning and control mechanisms). Second, we categorized different methods to include physical constraints in market models (considering branch flow equations and different tariffs). Third, we covered in detail different methods to calculate and allocate network tariffs and power losses within LEMs, which are possible due to digital advances of LEMs when comparing to traditional methods used in traditional networks setup. Finally, we have identified and listed several challenges that need to be addressed in relation to network impact of integration LEM models in distribution network.

**Acknowledgement:**

This publication is part of the work of the Global Observatory on Peer-to-Peer, Community Self-Consumption and Transactive Energy Models (GO-P2P), an Annex of the User-Centred Energy Systems Technology Collaboration Programme (Users TCP), under the auspices of the International Energy Agency (IEA). GO-P2P benefits from the support of Australia, Belgium, Ireland, Italy, The Netherlands, Switzerland, the United Kingdom and the United States.